\documentclass[12pt]{article}

\usepackage{amsmath,amsthm,amssymb,a4wide,amsfonts,latexsym}

\usepackage{mybbold}    %fuer die Eins
\usepackage[dvips]{graphicx}
\usepackage{subfigure}

\def\SU{\operatorname{SU}}

\newcommand{\ud}{{\rm d}}
\newcommand{\ui}{{\rm i}}
\newcommand{\ue}{{\rm e}}
\newcommand{\opH}{{\rm H}}
\newcommand{\opU}{{\rm U}}
\newcommand{\opK}{{\rm K}}
\newcommand{\opT}{{\rm T}}

\newcommand{\opR}{{\rm R}}
\newcommand{\opUt}{\opU_{\rm t}}
\newcommand{\opUs}{\opU_{\rm s}}

\newcommand{\vecB}{\boldsymbol{B}}
\newcommand{\vectB}{\tilde{\boldsymbol{B}}}

\newcommand{\vecz}{\boldsymbol{z}}

\newcommand{\vecr}{\boldsymbol{r}}

\newcommand{\vecgam}{\boldsymbol{\gamma}}
\newcommand{\vecsig}{\boldsymbol{\sigma}}

\newcommand{\vectau}{\boldsymbol{\tau}}
\newcommand{\vecrho}{\boldsymbol{\rho}}

\newcommand{\vecmu}{\boldsymbol{\mu}}

\newcommand{\vecpi}{\boldsymbol{\pi}}

\newcommand{\bY}{\boldsymbol{Y}}

\newcommand{\R}{{\mathbb R}}

\newcommand{\Z}{{\mathbb Z}}
\newcommand{\C}{{\mathbb C}}

\newcommand{\T}{{\mathbb T}}

\DeclareMathOperator{\uTr}{Tr}
\DeclareMathOperator{\utr}{tr}

\newcommand{\eins}{\mathmybb{1}}

\newcommand{\vecphi}{\boldsymbol{\phi}}

\newcommand{\abs}[1]{\left \lvert #1 \right \rvert}

\newcommand{\bB}{\boldsymbol{B}}
\newcommand{\bz}{\boldsymbol{z}}
\newcommand{\bzf}{\boldsymbol{z}_{\text{f}}}
\newcommand{\bm}{\boldsymbol{m}}
\newcommand{\bn}{\boldsymbol{n}}

\newcommand{\vp}[2]{\omega \left(#1, #2\right)}

\newcommand{\OpTH}{T_N}

\DeclareMathOperator{\Opp}{Op}
\newcommand{\Op}{\Opp_N}

\numberwithin{equation}{section}

\theoremstyle{definition}

\theoremstyle{remark}

\newcommand{\kommentar}[1]{}

\begin{document}

\thispagestyle{empty}

\noindent
ULM-TP/00-7 \\
December 2000\\

\vspace*{1cm}

\renewcommand{\thefootnote}{\fnsymbol{footnote}}

\begin{center}

{\LARGE\bf Quantum cat maps with spin 1/2\\}
\end{center}
\vspace{2ex}

\noindent
{\large\bf
Stefan Keppeler\footnotemark[1]%
\footnote[3]{E-mail address: {\tt kep@physik.uni-ulm.de}},
Jens Marklof\footnotemark[2]%
\footnote[4]{E-mail address: {\tt j.marklof@bristol.ac.uk}} and 
Francesco Mezzadri\footnotemark[2]%
\footnote[5]{E-mail address: {\tt f.mezzadri@bristol.ac.uk}}
}\\ 

\noindent
\footnotemark[1] 
Abteilung Theoretische Physik,
Universit\"at Ulm, 
Albert-Einstein-Allee 11,
D-89069 Ulm, Germany 

\noindent
\footnotemark[2]
School of Mathematics, 
University Walk,
University of Bristol, 
Bristol BS8 1TW, UK \\

\vfill

\noindent
{\bf Abstract:}
We derive a semiclassical trace formula for quantized 
chaotic transformations of the torus coupled to
a two-spinor precessing in a magnetic field.
The trace formula is applied to semiclassical correlation densities
of the quantum map, which, according to the conjecture
of Bohigas, Giannoni and Schmit, are expected to converge to
those of the circular symplectic  ensemble (CSE) of random matrices. 
In particular, we show that the diagonal approximation of
the spectral form factor for small arguments agrees with
the CSE prediction. The results are confirmed by numerical 
investigations.\\[2ex]
\noindent
{\small
PACS numbers: 05.45.Mt, 03.65.Sq\\ 
MSC numbers: 81Q50, 81Q20}

\vfill

\newpage

\section{Introduction}\label{intro}

Bohigas, Giannoni and Schmit \cite{BohGiaSch84} have conjectured that
the energy levels of quantized chaotic Hamiltonian systems
are distributed like eigenvalues of random matrices. 
The choice of the correct matrix ensemble depends on the behaviour
of the classical dynamics under time-reversal: the {\em unitary} ensemble (CUE)
for systems without time-reversal invariance, the {\em orthogonal}
or {\em symplectic} ensemble for systems with time-reversal invariance
with integer or half-integer spin, respectively.
Although there is much numerical evidence to support the conjecture
for systems following CUE and COE statistics, 
relatively few studies are concerned with systems
with half-integer spin 
\cite{CauGra89,SchDieKusHaaBer88,Sch89,TahBluSmi93}.

We shall here propose a simple model  
-- a quantized map of the torus coupled 
to a spin precessing in a magnetic field -- 
which is easily accessible both from the numerical and the analytical point of
view: Suppose our particle's initial position is $q_0$ with momentum $p_0$.
With $(p_0,q_0)$ fixed, we allow the spin to precess for 
one time unit 
in the magnetic field $\vecB(q_0)$; 
we then apply the Anosov map 
\begin{equation}
\vecphi: \T^2\rightarrow \T^2, \qquad 
\begin{pmatrix} p_0 \\ q_0 \end{pmatrix} \mapsto  
\begin{pmatrix} p_1 \\ q_1 \end{pmatrix} \mod 1,
\end{equation}
keeping the spin  fixed; thereafter the spin
is allowed to precess again for one time unit  
in $\vecB(q_1)$, and so on. 

The idea here is  that, due to the chaotic
dynamics of $\vecphi$, the values of the magnetic field at the iterated
positions $q_0,q_1,q_2,\ldots$ are essentially uncorrelated 
and the spin precession therefore behaves like a random walk in $\SU(2)$.
The ergodic properties of such constructions,
so-called {\em skew products}, 
are well understood \cite{Bri75,KeyNew74,KeyNew76,Noo97}.

The quantization of our model in terms of unitary $2N\times 2N$
matrices $\opU$ is straightforward (section \ref{model}). 
In section \ref{symmetry}
we discuss in detail those anti-unitary operators $\opT$
which correspond to time-reversal symmetries in that
$\opT^{-1} \opU \opT=\opU^{-1}$. If $\opT$ is the only
such symmetry present, and if in addition  $\opT^2=-\eins_{2N}$,
one expects the spacing distribution for the eigenvalues of $\opU$
to converge in the  semiclassical limit $N\rightarrow\infty$
to that of the symplectic ensemble CSE.

Indeed, for generic Anosov maps $\vecphi$ we numerically
observe a transition from
COE to CSE statistics when the magnetic field is switched on
(section \ref{numerics}).
It is, however, remarkable  that even unperturbed cat maps,
whose spectrum is highly degenerate \cite{KeaMez00,KulRud00},
exhibit CSE statistics for non-zero magnetic fields.

The remainder of our paper is devoted to the semiclassical analysis
of spectral correlation densities. 
In section \ref{stf} we derive a semiclassical trace formula
which expresses sums over eigenvalues of $\opU$ in terms of 
 sums over fixed points of the classical Anosov map $\vecphi$.  
We observe that the fixed points remain unchanged when switching on the 
magnetic field but that additional weight factors, taking care of the spin 
contribution, appear. In the case of flows an analogous observation 
has been made in \cite{BolKep98,BolKep99a}. 
In section \ref{formf} the trace formula is employed to investigate
two-point correlations of the eigenvalues of $\opU$,
with focus on the behaviour of the spectral form factor 
%i.e., the Fourier transform of the two-point correlation density,
for small arguments. In particular, we show that   the 
diagonal approximation \cite{HanOzo84,Ber85}
reproduces the random matrix prediction
if the magnetic field is sufficiently generic. 
We essentially follow the line of arguments in \cite{BolKep99b},
where the form factor for continuous-time dynamics is discussed.
 However, the assumption in \cite{BolKep99b} that the 
skew product dynamics must be mixing is replaced by a simpler condition. 
In the final section \ref{piecewise} we discuss the interesting 
fact that the diagonal approximation also  works if  the 
spin precession does not become equidistributed in $\SU(2)$,
but only takes a finite number of distinct values in $\SU(2)$.

\section{Maps with spin}\label{model}

The Hamiltonian of a particle with spin $\frac12$ in
a magnetic field $\vecB(q)$ is given by (in units
where $2mc/e=1$)
\begin{equation}
  \opH = \opH_0 \eins_2 - \hbar\, \vecsig\cdot \vecB(q) ,
\end{equation}
where $\opH_0$ is a scalar Hamiltonian, 
and the second term describes the interaction between
the magnetic moment of spin and the magnetic field $\vecB(q)$. 
Here $\eins_2$ denotes the $2 \times 2$ unit matrix, and
$\vecsig$ is the vector of Pauli matrices
\begin{equation}
\sigma_x=\begin{pmatrix} 0 & 1 \\ 1 & 0 \end{pmatrix}, \quad 
\sigma_y=\begin{pmatrix} 0 & -\ui \\ \ui & 0 \end{pmatrix}, \quad 
\sigma_z=\begin{pmatrix} 1 & 0 \\ 0 & -1 \end{pmatrix}.
\end{equation}

We are interested in modeling free spin precession interrupted
only by periodic delta-shaped kicks at integer times $t=n\in\Z$.
Such a system is represented by the  time dependent Hamiltonian 
\begin{equation}
  \opH(t) = 
  \sum_{n\in\Z} \opH_0 \eins_2 \, \delta(t-n) 
  - \hbar \, \vecsig\cdot \vecB(q)  ,
\end{equation}
cf.~\cite{Sch89,TahBluSmi93}.
The Floquet operator which propagates the system, e.g., from time 
$t=-\frac12$ to $t=\frac12$, is then given by
\begin{equation}
\label{defsys}
  \opU = \opUs \, \opUt \, \opUs
  \, , \quad \text{with} \quad 
  \opUs = \ue^{\frac{\ui}{2}\vecsig \cdot\vecB(q)}
  \quad \text{and} \quad 
  \opUt=\ue^{-\frac{\ui}{\hbar} \opH_0}\eins_2 . 
\end{equation}

In this paper, the translational dynamics $\opUt$ will be represented by 
a quantized map of the torus $\T^2$, which is given by a unitary 
operator $\opU_N$ acting on an $N$ dimensional Hilbert space
${\mathcal H}_N \simeq \C^N$. The dimension of the Hilbert 
space ${\mathcal H}_N$
and Planck's constant are related via the condition $2\pi\hbar N = 1$. 
 (We refer the reader to appendix A 
for more details on the quantization of maps
on the torus.)  For instance, in the case of the  map  
\begin{equation}
\label{cm}
A: \begin{pmatrix}
p \\ q 
\end{pmatrix}
\mapsto
\begin{pmatrix}
2 & 3 \\ 1 & 2 
\end{pmatrix}
\begin{pmatrix}
p \\ q 
\end{pmatrix}
=
\begin{pmatrix}
2p+3q \\ p+ 2q 
\end{pmatrix} \mod 1 ,
\end{equation}
the action of $U_N$  on functions  $\Phi\in \mathcal{H}_N$
is  given by \cite{HanBer80}
\begin{equation}
\label{fopj}
[\opU_N(A) \Phi](Q)= 
\left(\frac{1}{\ui N}\right)^{1/2}
\sum_{Q'=0}^{N-1} \exp\left[\frac{2 \pi \ui}%
    {N}\left(Q^2 -QQ' + {Q'}^2\right)\right] \Phi(Q')
\end{equation}
$Q=0,\ldots, N-1$. We shall also investigate the
perturbed cat map 
\begin{equation}
\vecphi : \vecrho \circ A \circ \vecrho
\end{equation}
with perturbation
\begin{equation}
\label{pcm}
\vecrho : \begin{pmatrix}
p \\ q 
\end{pmatrix}
\mapsto
\begin{pmatrix}
p + \frac{k}{2} f(q) \\ q 
\end{pmatrix} \mod 1 ,
\end{equation}
where $f$ is a bounded periodic function and $k$ measures the strength of
perturbation. Since Anosov systems are structurally stable, $\vecphi$ 
remains Anosov if $k$ is small enough.
The quantization of $\vecphi$ is represented by \cite{BdMOdA95}
\begin{equation}
\label{fopj2}
\opU_N (\vecphi) = \opU_N(\vecrho)\,\opU_N (A)\,\opU_N (\vecrho)
\end{equation}
with
\begin{equation}
[\opU_N (\vecrho) \Phi](Q)= \ue^{\pi \ui N k S\left(\frac{Q}{N}\right)}
\Phi(Q)
\end{equation}
and $f(q)= \frac{\ud}{\ud q}S(q)$.

The propagators $\opUt(A)$ and $\opUt(\vecphi)$ now act on a two-spinor 
$\Psi\in \C^2 \otimes \mathcal{H}_N $ simply by
\begin{equation}
\label{propf}
\opUt(A) \Psi = \begin{pmatrix} \opU_N (A)\Phi_1 \\ \opU_N(A) \Phi_2
\end{pmatrix}, \qquad
\opUt(\vecphi) \Psi = \begin{pmatrix} \opU_N (\vecphi)
\Phi_1 \\ \opU_N(\vecphi) \Phi_2
\end{pmatrix}, \qquad
\Psi = \begin{pmatrix} \Phi_1 \\ \Phi_2 \end{pmatrix},
\qquad  \Phi_1, \Phi_2 \in \mathcal{H}_N ,
\end{equation}
 and the action of $\opUs$ is straightforwardly given by
\begin{equation}\label{def_sigxB}
  [\opUs \Psi](Q) = \ue^{\frac{\ui}{2} \vecsig\cdot \vecB 
\left(\frac{Q}{N}\right)}
  \Psi(Q) .
\end{equation}

\section{Anti-unitary symmetries}
\label{symmetry}

Following Bohigas, Giannoni and Schmit \cite{BohGiaSch84}, we expect
that the eigenvalue statistics of the quantized map $\opU$ converge
to those of the circular random matrix ensembles, CUE, COE, or CSE.
 The choice of the correct ensemble depends on whether the
quantum map possesses an anti-unitary  symmetry $\opT$.  
In particular, the
spectral fluctuations are expected to agree with those of the CSE
ensemble if $\opT^2 =-\eins_{2N}$. In this section we discuss the 
necessary conditions to be imposed on the classical map $\vecphi$
and on the magnetic field $\vecB(q)$ in order to observe CSE
statistics. 

To this end consider an Anosov map $\vecphi$ invariant under time
reversal, i.e.
\begin{equation} \label{timerev}
\vectau\circ\vecphi\circ\vectau = \vecphi^{-1} , 
\quad  {\rm where} \quad 
\vectau: 
\begin{pmatrix}
p \\ q 
\end{pmatrix}
\mapsto
\begin{pmatrix}
-p \\ q 
\end{pmatrix}.
\end{equation}
Moreover, suppose that the map $\vecphi$ is also invariant under
inversion, i.e.
\begin{equation}\label{inversion}        
\vecpi \circ \vecphi \circ \vecpi = \vecphi \quad \text{with} \quad 
\vecpi : \begin{pmatrix}
p \\ q 
\end{pmatrix}
\mapsto
\begin{pmatrix}
-p \\ -q 
\end{pmatrix}.
\end{equation}
 Then, $\vecphi$ will also be invariant under a transformation with 
\begin{equation}\label{noncontr}
\tilde{\vectau} = \vecpi \circ \vectau ,
\end{equation} 
which we will call {\em non-conventional} time-reversal, see, 
e.g. \cite{Haa91} for a general discussion.
For spinless motion 
the quantum time reversal operator $\opK$ is simply complex conjugation,

\begin{equation}
[\opK\Psi](Q)=\overline\Psi(Q) ,
\end{equation}
and the operator corresponding to $\tilde{\vectau}$ is  given by
\begin{equation}
[\tilde{\opK}\Psi](Q)=\overline\Psi(-Q) .
\end{equation}
 Moreover, since $\vectau$ and $\tilde{\vectau}$ are symmetries of
$\vecphi$, the quantum map $\opUt(\vecphi)$ will be invariant under
$\opK$ and $\tilde{\opK}$, i.e.
\begin{equation}
  \opK \, \opUt(\vecphi) \, \opK \,   
= \opUt^{-1}(\vecphi) , \qquad 
  \tilde{\opK} \, \opUt(\vecphi) \, \tilde{\opK} \,   
= \opUt^{-1}(\vecphi).
\end{equation}
In general, however,  the coupled map $\opU(\vecphi)=\opUs
\opUt(\vecphi) \opUs$ is not invariant under these symmetries, because
\begin{subequations}
\begin{gather}   
\label{btq}
  \left[ \opK \,  \opUs \, \opK \,
  \Psi  \right](Q)  = \ue^{- \frac{\ui}{2} \vecsig\cdot \vectB(\frac{Q}{N})} 
  \, \Psi(Q) , \\
\label{btq2}
 \left[ \tilde{\opK} \,  
    \opUs  \, \tilde{\opK} \,
  \Psi  \right](Q) = \ue^{- \frac{\ui}{2} \vecsig\cdot \vectB(-\frac{Q}{N})} 
  \, \Psi(Q) ,
\end{gather}
\end{subequations}
where $\vectB(q)=(B_x(q),-B_y(q),B_z(q))$. 
Since we are  concerned with system with spin, we should also 
invert the direction of spin when reversing time. 
Let us consequently define the modified time-reversal operators $\opT$
and $\tilde{\opT}$ by
\begin{equation}\label{spintr}
[\opT\Psi](Q)= \ue^{\ui \pi \sigma_y/2} \, \overline\Psi(Q) 
             = \ui\sigma_y \overline\Psi(Q) , \qquad
[\tilde{\opT}\Psi](Q)=\ui\sigma_y \overline\Psi(-Q) .
\end{equation}
Note that $\opT^2=\tilde{\opT}^2=-\eins_{2N}$,  
as needed for CSE statistics.  For these operators one obtains
\begin{subequations}
\begin{gather}   \label{sym1}
  \left[ \opT^{-1} \, \opUs  \, \opT
  \, \Psi \right](Q)= \ue^{\frac{\ui}{2} 
\vecsig\cdot \vecB(\frac{Q}{N})} \Psi(Q) , \\
\left[ \tilde{\opT}^{-1} \, \opUs \, \tilde{\opT} \label{sym2}
  \, \Psi \right](Q)= \ue^{\frac{\ui}{2} 
\vecsig\cdot \vecB(-\frac{Q}{N})} \Psi(Q) .
  \end{gather}     
\end{subequations}  
From (\ref{sym1}) we easily see that 
$\opT^{-1} \, \opUs  \, \opT \neq \opUs^{-1}$ for non-zero
magnetic fields, since obviously 
$\vecB(q)\neq -\vecB(q)$. However, rel.~(\ref{sym2}) implies 
that $\tilde{\opT}^{-1} \, \opUs  \, \tilde{\opT} = \opUs^{-1}$
for odd magnetic field, i.e. if $\vecB(-q) = -\vecB(q)$.

 In addition we have to make sure that there is no further anti-unitary
symmetry, in particular the system must neither be invariant under 
$\opK$ nor under $\tilde{\opK}$.  
If $B_y(q)$ vanishes identically then
$\vectB(q)=\vecB(q)$ and by \eqref{btq} the spin part is also 
invariant under $\opK$, which should be avoided.
 The last argument is of course independent of the choice
of the coordinate system, and hence we require that 
{\em there is no vector $\vecr\neq0$ such that $\vecr\cdot\vecB(q)=0$
for all $q$}, i.e. the magnetic field $\vecB(q)$ must have linearly 
independent components. 

 To summarize, in order to observe CSE spectral distributions
we shall require that $\tilde{\vectau}$ is a symmetry of the classical
map $\vecphi$, that the magnetic field $\vecB(q)$ is odd and
that its components are independent functions. 

\section{Spectral statistics}
\label{numerics}

 After having clarified under which 
 conditions we  expect
to see CSE statistics we will now test our arguments in a numerical 
experiment for the quantized cat map $A$ and its perturbation
$\vecphi$.  A simple choice of the shear  in \eqref{pcm} 
is, e.g., given by
\begin{equation}
\label{pc}
f(q)= \frac{1}{2\pi} \sin 2 \pi q.
\end{equation}
 The map  $\vecphi$ will be Anosov if \cite{BdMOdA95}
\begin{equation}
  k < k_{\text{max}}= \frac{\sqrt{3}-1}{\sqrt{5}}=0.327...
\end{equation}
  The only symmetries of the cat map~\eqref{cm} 
are time-reversal $\vectau$ \eqref{timerev}, 
inversion $\vecpi$ \eqref{inversion}, and 
non-conventional time-reversal $\tilde{\vectau}$ \eqref{noncontr}. 
The choice of the perturbation~\eqref{pc}, with $f$ being odd,
conserves all symmetries $\vectau$, $\tilde{\vectau}$ and $\vecpi$.

Now, in order to observe CSE
statistics, the spin precession must be caused by a magnetic field
 which is odd and periodic in $q$ and whose
components are independent functions (cf. section~\ref{symmetry}).  
These constraints guarantee
that the only anti-unitary  symmetry of the quantum 
system is $\tilde{\opT}$ \eqref{spintr}.  
For our numerical calculation we chose the magnetic field 
\begin{equation}
\label{magf}
\bB(q)= \begin{pmatrix} \sin (2 \pi q)\\
           \sin (4 \pi q) \\ \sin (6 \pi q) \end{pmatrix}, 
\end{equation}
which clearly satisfies these requirements.

We now combine $\opU_N(A)$ and
$\opU_N(\vecphi)$ with the spin precession propagator $\opUs$ as
in~\eqref{defsys}  to
obtain the quantum maps
\begin{equation}
\label{maptd}
\opU(A) = \opUs \opUt(A) \opUs \quad \text{and} \quad \opU(\vecphi)=
\opUs \opUt(\vecphi) \opUs,
\end{equation}
where
\begin{equation}
\opUt(A)=\begin{pmatrix} \opU_N(A) & 0 \\ 0 & \opU_N(A) \end{pmatrix}
\quad \text{and} \quad 
\opUt(\vecphi)=\begin{pmatrix} \opU_N(\vecphi) & 0 \\ 0 & \opU_N(\vecphi) 
\end{pmatrix}.
\end{equation}
The matrix elements of $\opU(A)$ and $\opU(\vecphi)$ are easily
worked out using~\eqref{fopj} -- \eqref{def_sigxB}.  
 If $N$ is divisible by four, there exist simple relations between
the matrix elements of the propagators in eqs.~\eqref{fopj}
and~\eqref{fopj2}~\cite{BerKeaPra98} which lead to invariance with respect to
the unitary operator  
\begin{equation}
 \left[\opR\Psi \right](Q)=\ui \ue^{\ui \pi Q}\Psi(Q + N/2).
\end{equation}
This operator obeys the relations
\begin{equation}
[\opU(A),\opR]=0, \quad  [\opU(\vecphi), \opR]=0,
\quad [\opR,\tilde{\opT}]=0  \quad \text{and}
\quad \opR^2 = - \eins_{2N}. 
\end{equation}
As a consequence in this case the statistics of the spectra of
$\opU(A)$ and $\opU(\vecphi)$ would follow those of the CUE ensemble, cf. 
\cite{Haa91}.  Thus, we shall require that $ N \not \equiv 0 \bmod 4$.

Studying the spectral statistics of quantum maps means looking at the
distributions of the eigenangles $\{\alpha_1,\ldots,\alpha_{2N} \}$.
In order to compare energy fluctuations in different systems, the
spectra must be unfolded, that is the energy levels are rescaled so
that the mean level spacing is one. (The spacing $s$ is defined
as the distance between two consecutive levels.) 
Moreover, in order to observe universal distributions
the spectra must be desymmetrized, that is only energy levels with
the same quantum numbers corresponding to all mutually commuting 
unitary symmetries  must be considered.  Therefore, when 
analyzing the spectra of $\opU_N(A)$ and $\opU(\vecphi)$ in the 
absence of spin dynamics, we have to take into account only eigenphases
which correspond to the same eingenvalue $P=\pm 1$ of $\vecpi$. 
For systems with half-integer
spin and time-reversal invariance  an additional proviso must be taken
into account: each energy level  has at least multiplicity two 
 (Kramers'  degeneracy \cite{Kra30})
because the square of time-reversal 
operators $\opT$ for systems with half-integer spin being $-\eins_{2N}$
causes $\opT \Psi$ to be orthogonal to $\Psi$  \cite{Wig32}. 
The statistical analysis, however, must 
be performed on a spectrum obtained by removing such degeneracy.
Compare section \ref{formf} for details.  

\begin{figure}
\centering \subfigure[Spacing distribution of the quantum 
cat map~\eqref{fopj} with $P=+1$.]{\label{sdcm}
  \includegraphics[width=2.9in]{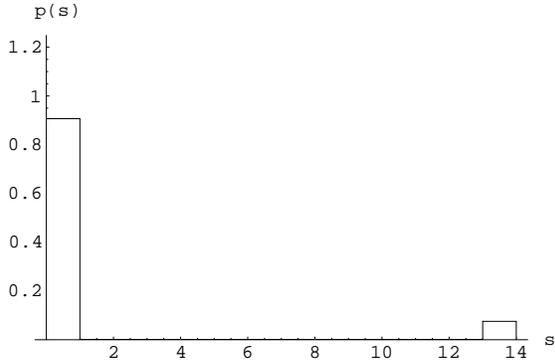}} \hspace{.2in}
\subfigure[Cumulative spacing distribution of the 
quantum cat map~\eqref{fopj} with $P=+1$.]{\label{csdcm}
  \includegraphics[width=2.9in]{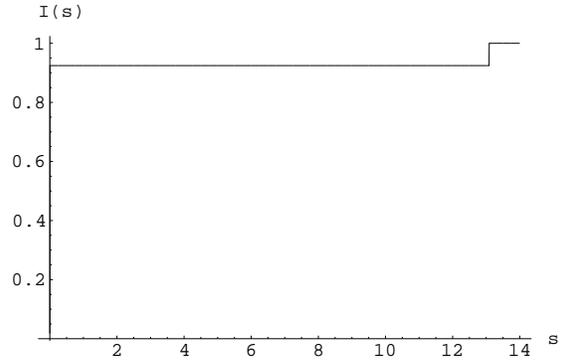}}\\
\subfigure[Spacing distribution of the quantum map~\eqref{fopj2} with
k=0.32 and $P=+1$.]{ \includegraphics[width=2.9in]{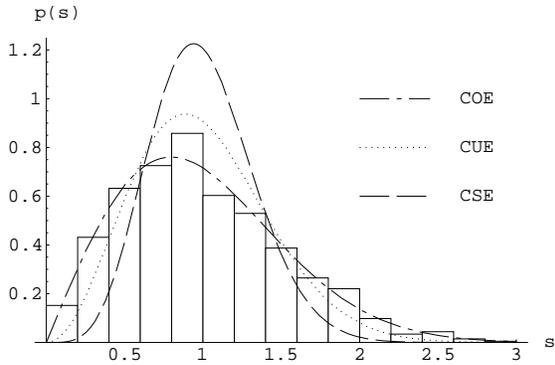}}
\hspace{.2in} \subfigure[Cumulative spacing distribution of the
quantum map~\eqref{fopj2}with k=0.32 and $P=+1$.]{
  \includegraphics[width=2.9in]{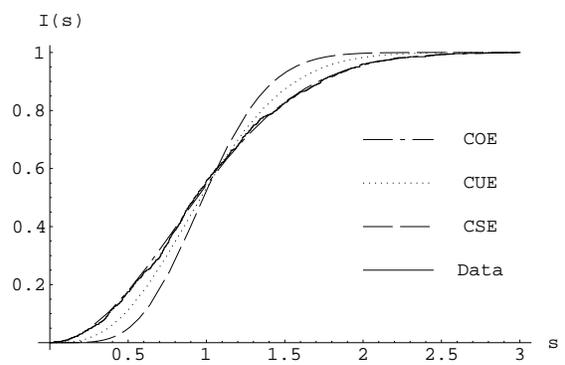}}
      \caption{Local spectral statistics  of the quantum cat
        map~\eqref{fopj} (above)
        and of its perturbation~\eqref{fopj2} (below).
        The eigenangles belong to states with the same eigenvalue of 
        $\vecpi$ and their number is $N=1021$.}
        % The RMT curves are the Wigner surmises.
      \label{fig1}
\end{figure}

Let us briefly describe the spectral distributions of $\opU_N(A)$ and
$\opU_N(\vecphi)$ in the absence of  spin dynamics.  
The spectra of quantum cat maps $\opU_N(A)$ are well known 
to be highly non-generic, and in particular do not follow
any universal distribution  \cite{Kea91a}. 
The reason for this untypical behaviour is
the high number of quantum symmetries -- about $O(N)$
 many~\cite{KeaMez00,KulRud00} --
which commute with the quantum propagator $\opU_N(A)$.  
Such arithmetical symmetries  can easily be broken 
by slightly perturbing the classical map~\cite{BdMOdA95}, 
and the expected  COE random matrix statistics are recovered.  This
behaviour is evident in fig.~\ref{fig1}, 
where the consecutive level spacing distribution $p(s)$  
and the integrated level spacing distribution  
\begin{equation}
I(s)=\int_0^sp(s') \, \ud s'
\end{equation}
are plotted  for the quantum maps~\eqref{fopj} 
and~\eqref{fopj2}. 

\begin{figure}
\centering \subfigure[Spacing distribution of the coupled map  $\opU(A)$.]{
  \includegraphics[width=2.9in]{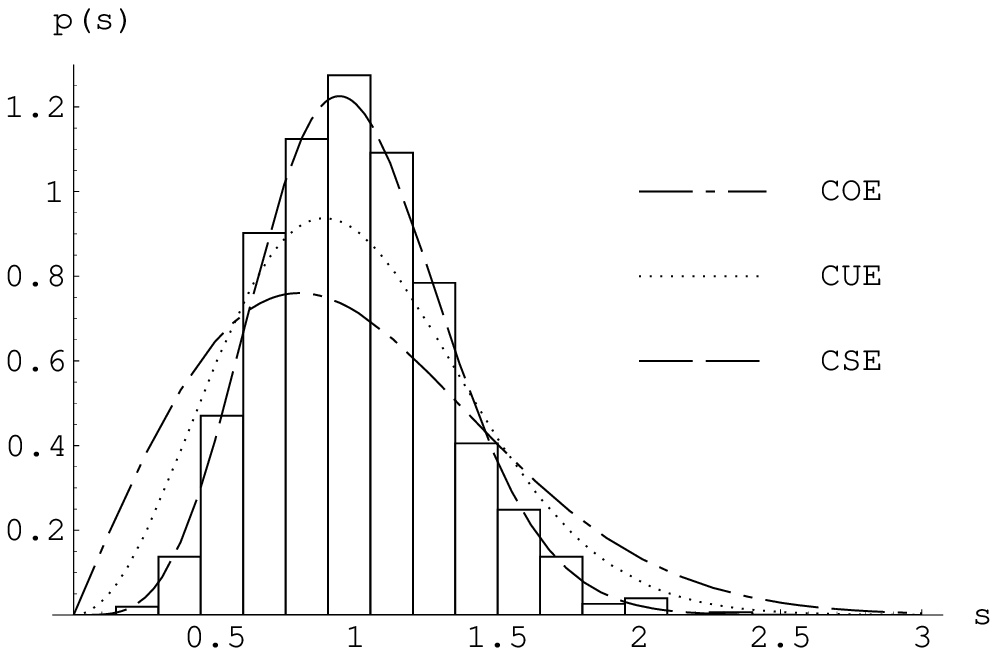}} \hspace{.2in}
\subfigure[Cumulative spacing distribution of the coupled map $\opU(A)$.]{
  \includegraphics[width=2.9in]{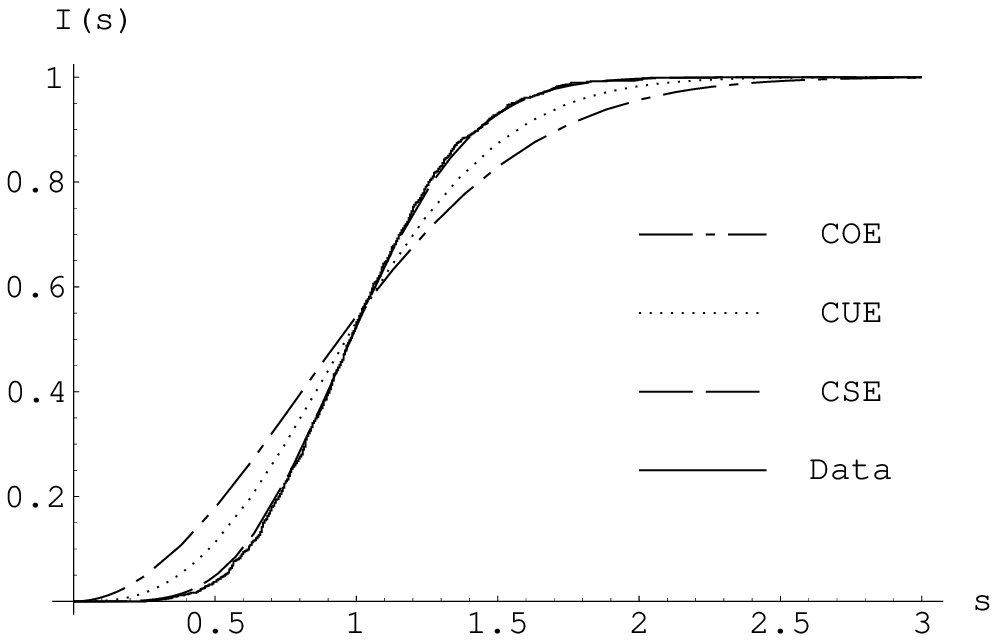}}\\
\subfigure[Spacing distribution of the coupled map 
 $\opU(\vecphi)$ with k=0.32.]{
  \includegraphics[width=2.9in]{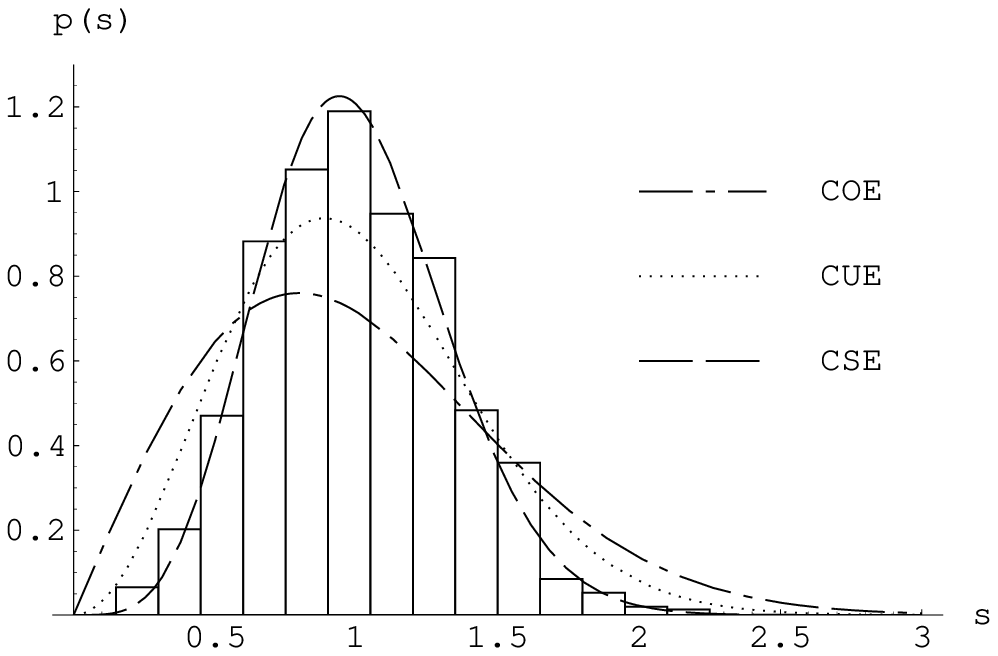}} \hspace{.2in}
\subfigure[Cumulative spacing distribution of the coupled map 
 $\opU(\vecphi)$ with
k=0.32.]{ \includegraphics[width=2.9in]{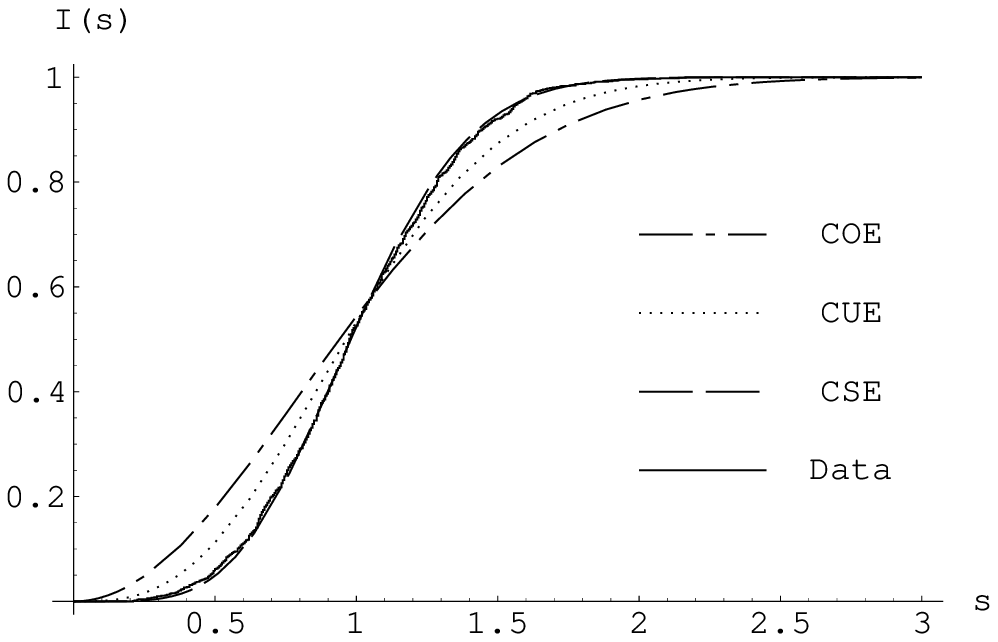}}
      \caption{Local spectral statistics  of the quantum cat
        map (above) and of its perturbation (below) when the spin
        dynamics is introduced.  The number of eigenangles $N$ is
        $1021$.}
        % The RMT curves are the Wigner surmises.
        \label{fig2}
\end{figure}

Fig.~\ref{fig2} shows the spectral statistics of $\opU(A)$ and
$\opU(\vecphi)$ defined in~\eqref{maptd} with the
magnetic field given by~\eqref{magf}.
Clearly, in both cases the spectral distributions nicely follow 
CSE statistics, as predicted.  It is  remarkable that the non-generic
behaviour of the quantum cat map without spin (figs.~\ref{sdcm}
and~\ref{csdcm}) has disappeared, even though the spin coupling
is of sub-principal order in $\hbar$ and therefore does not affect
the highly degenerate classical dynamics. 
 The semiclassical arguments
in the following sections will 
explain the CSE statistics for generic maps $\vecphi$, but in the case
of the unperturbed cat map $A$ the situation is more subtle. We will 
briefly  return to this point in section \ref{piecewise} when 
discussing piecewise constant magnetic fields.

\section{Semiclassical trace formula}
\label{stf}

The purpose of this section is to derive a trace formula for the
quantum maps~\eqref{maptd}. 
Since Berry's study of  the spectral rigidity \cite{Ber85} based 
on Gutzwiller's trace formula \cite{Gut71}, trace formulae are 
the main semiclassical tools  in analyzing spectral correlations.
In the case of flows trace formulae including spin contributions have 
been derived in \cite{BolKep99a} for the Pauli and the Dirac equation.

In what follows $\vecphi$ can be any Anosov map on  $\T^2$.  
The magnetic field $\vecB$ will
be left generic too, except that, for simplicity, it should be a
function of $q$ only.  Our goal is to determine
$\uTr[\opU^n(\vecphi)]$ as a sum over periodic orbits. 
For this purpose it is  more
convenient to consider the operator
\begin{equation}
\tilde{\opU}(\vecphi)=\opUt(\vecphi)\opUs^2.
\end{equation}
Since $\tilde{\opU}(\vecphi)$ is unitarily conjugate to
$\opU(\vecphi)$, clearly $\uTr[\opU^n(\vecphi)]=
\uTr[\tilde{\opU}^n(\vecphi)]$.

Let $\Op(f)$ denote the Weyl quantization of a classical observable $f
\in C^{\infty}(\T^2)$ (see appendix~\ref{qmt}).  For a $2\times 2$ matrix 
\begin{equation}
g(\vecz) = \begin{pmatrix} g_{11}(\vecz) & g_{12}(\vecz) 
                        \\ g_{21}(\vecz) & g_{22}(\vecz)
         \end{pmatrix},  \quad \vecz = (p,q) \in \T^2 , 
\end{equation}
whose elements are smooth functions on $\T^2$, 
we then define
\begin{equation}
\Op(g) = \begin{pmatrix} \Op(g_{11}) & \Op(g_{12}) \\
                         \Op(g_{21}) & \Op(g_{22})
         \end{pmatrix}, 
\end{equation}
which semiclassically fulfills 
(cf.~appendix~\ref{qmt})
\begin{equation}\label{sc_factor}
\Op(g_k g_{k-1}\ldots g_1) \sim \Op(g_k)\Op(g_{k-1}) \ldots \Op(g_1), \quad N
\rightarrow \infty.
\end{equation}
 With the choice 
\begin{equation}
g(\vecz)= \exp\left(\ui \vecsig\cdot \vecB(q)\right) \in {\rm SU}(2) 
\end{equation}
we have $\opUs^2=\Op(g)$ and, thus, we may write 
$\tilde{\opU}(\vecphi)=\opUt(\vecphi)\Op(g)$.
We now rearrange $\tilde{\opU}^n(\vecphi)$ as follows:

\begin{multline}
\label{trr}
\tilde{\opU}^n(\vecphi) = 
\opUt^n (\vecphi) 
\left( \opUt^{1 -n}(\vecphi) \Op(g) \opUt^{n-1}(\vecphi) \right)  
\left( \opUt^{2 -n}(\vecphi) \Op(g)\opUt^{n-2}(\vecphi) \right) \cdots 
\\ 
\cdots \left( \opUt^{-1}(\vecphi) \Op(g)\opUt(\vecphi) \right) \Op(g).
\end{multline}
Applying Egorov's theorem (appendix~\ref{qmt}) 
\begin{equation}
\opUt^{k-n}(\vecphi)\Op(g)\opUt^{n-k}(\vecphi)
\sim \Op(g \circ \phi^{n-k})  , \quad N \to \infty
\end{equation}
to each factor ($k=1,\ldots,n$)  yields
\begin{equation}
\label{utild}
\tilde{\opU}^n (\vecphi)\sim \opUt^n (\vecphi)
\prod_{k=0}^{n-1} \Op(g \circ \vecphi^k) , \qquad
N \rightarrow \infty
\end{equation} 
where the product is time ordered, i.e.
\begin{equation}
\prod_{k=0}^{n-1} \Op(g \circ \vecphi^k) = \Op(g \circ
\vecphi^{n-1})\Op(g \circ \vecphi^{n-2})\ldots \Op(g).
\end{equation}
In order to simplify the notation, it is convenient to set
\begin{equation}
g_k(\bz) = g \circ \vecphi^{k-1}(\bz), \quad  k=0, \ldots, n,
\end{equation}
with the convention that $g_0(\bz)= \eins_2$, and
$g_1(\bz)=g(\bz) = \exp \left(\ui \vecsig\cdot \vecB(q) \right)$.
Now eq.~\eqref{utild} becomes
\begin{equation}
\label{neq}
\tilde{\opU}^n (\vecphi)\sim 
\opUt^n(\vecphi)\prod_{k=0}^{n} \Op(g_k), \quad N\rightarrow \infty.
\end{equation}
In the case when $\vecphi=A$ is an unperturbed cat map, Egorov's
theorem is an identity, not  merely an asymptotic relation, and  
hence eq.~\eqref{neq} is an identity too. Due to (\ref{sc_factor})  in 
the semiclassical limit ($N \rightarrow \infty$) eq.~\eqref{neq} becomes
\begin{equation}
\label{neq2}
\tilde{\opU}^n(\vecphi) \sim 
\opUt^n (\vecphi) \Op(d_n), 
\end{equation}
where
\begin{equation}
\label{defd}
d_n(\bz) = \prod_{k=0}^{n}g_k(\bz) \in {\rm SU}(2). 
\end{equation}
Let $\varphi_1,\ldots,\varphi_N$ be an orthogonal basis of
eigenfunctions of $\opU_N(\vecphi)$ with eigenphases $\omega_j$,
\begin{equation}
\opU_N(\vecphi) \varphi_j=\ue^{\ui \omega_j}\varphi_j .
\end{equation}
Taking the  trace of the right-hand side of~\eqref{neq2} yields
\begin{equation}
\label{trs}
\uTr \left[\opUt^n (\vecphi) \Op(d_n)\right]
= \sum_{j=1}^N (\varphi_j, \Op(\utr d_n) \varphi_j)\, \ue^{\ui n \omega_j}.
\end{equation}
The leading order term in the semiclassical limit $N\rightarrow\infty$ of 
the above can be shown to be  (cf.~appendix \ref{trss})
\begin{equation}\label{matrform}
\sum_{j=1}^N (\varphi_j, \Op(\utr d_n) \varphi_j)\, \ue^{\ui n \omega_j}
\sim
\sum_{\bzf}
 \frac{\utr d_n(\bzf)}{\sqrt{-R^{\left(n\right)}_{\bzf}}}
\exp\left[2 \pi \ui N\left( S_{\vecphi^n}
(q_{\text{f}} + m_1,q_{\text{f}}) - m_2q_{\text{f}}\right)
\right] .
\end{equation}
Here $\bzf$ are fixed points of order $n$ with winding numbers 
$(m_1,m_2)=\bm $, i.e. 
\begin{equation}
\vecphi^n(\bzf)=\bzf + \bm.
\end{equation}
 Moreover, $R^{\left(n\right)}_{\bzf}= \det
(\mathcal{M}^n_{\bzf}- I)$, where 
$\mathcal{M}^n_{\bzf}:= \frac{\ud\vecphi^n}{\ud\vecz}(\bzf)$ is the
monodromy matrix and $S_{\vecphi^n}$ 
is the generating function of $\vecphi^n$ on the covering plane.
We will denote the classical action of the fixed point $\bzf$ by 
\begin{equation}
\label{action}
S_{\vecphi^n}(\bzf):=
S_{\vecphi^n}(q_{\text{f}} + m_1,q_{\text{f}}) - m_2q_{\text{f}} .
\end{equation}
Thus, in leading semiclassical order ($N\rightarrow\infty$) we find
\begin{equation}
\label{trform}
\uTr\left[\opU^n \left(\vecphi \right)\right]  \sim
\sum_{\bzf}
 \frac{\utr d_n(\bzf)}{\sqrt{-R^{\left(n\right)}_{\bzf}}}
\exp\left[2 \pi \ui N  S_{\vecphi^n}(\bzf) \right] ,
\end{equation}
which is the main result of this section.
Note that~\eqref{trform} is always 
a semiclassical expression,
even if  $\vecphi=A$ is a cat map.  Semiclassical approximations enter
in~\eqref{neq2} and in~\eqref{matrform} for both the unperturbed and
perturbed map.  
Formula~\eqref{trform} shows that the classical orbits are not
affected by spin precession  and that the spin contribution is
represented by the factors $\utr d_n(\bzf)$ for each periodic orbit. 
The respective result in the case of flows was obtained in 
 \cite{BolKep98,BolKep99a}. 

\section{Semiclassical analysis of two-point correlations}
\label{formf}

We will now apply the trace formula derived in the previous
section to the study of spectral two-point correlations, 
essentially following the line of arguments of \cite{BolKep99b}, 
giving a new argument concerning equidistribution in ${\rm SU}(2)$.

Since $\opU$ is invariant under $\tilde{\opT}$ with
$\tilde{\opT}^2=-\eins_{2N}$,
cf. section \ref{symmetry}, it is easily seen that the
eigenvalues must have at least multiplicity two (Kramers'  degeneracy). 
 Let us denote by $\alpha_1,\ldots,\alpha_{2N}\in[0,2\pi)$
the eigenphases of $\opU$ with eigenfunctions $\Psi_j$, i.e. 
\begin{equation}
\opU \Psi_j = \ue^{\ui \alpha_j} \Psi_j , 
\end{equation}
  labelled such that $\alpha_k=\alpha_{k+N}$,
for $k=1,\ldots, N$. We will then only need to consider 
correlations between the first 
$N$ distinct eigenphases $\alpha_1,\ldots,\alpha_N$.

In order to measure correlations on the scale of the mean level
spacing, we rescale the spectrum by putting
\begin{equation}
  x_k = \frac{N}{2\pi} \, \alpha_k  ,\qquad k=1,\ldots, N.
\end{equation}
The two-point correlation density of this sequence is defined as 
\begin{equation}
\label{twop}
  R_2(s,N)= \frac{1}{N}\sum_{k,l=1}^{N}\sum_{m\in\Z}
  \delta\left(s -(x_k - x_l)- Nm\right) -1  ,
\end{equation}
and its Fourier transform,
the {\em spectral form factor} $K_2(\tau,N)$, reads
\begin{equation}
\begin{split}
  K_2(\tau,N) &= \int_0^{N} R_2(s,N) \, \ue^{2\pi\ui\tau s} \, \ud s
  = \frac{1}{N} \left| \sum_{k=1}^{N} \ue^{2\pi\ui x_k \tau} \right|^2
    - N \delta_{n0} \\
  &= \frac{1}{4N} \left| \uTr (\opU^n) \right|^2 - N \delta_{n0} \, .
\end{split}
\end{equation}
for $\tau=\frac{n}{N}$, $n\in\Z$.
According to the Bohigas-Giannoni-Schmit conjecture \cite{BohGiaSch84}, 
one  expects that semiclassically 
($N\rightarrow\infty$) the form factor
$K_2(\tau,N)$ converges on average to the corresponding CSE form factor.
That is, for any smooth and rapidly decaying test function $\phi$,
 we expect
\begin{equation}       \label{theBGS}
  \lim_{N \to \infty} \sum_{n\in\Z} K_2 \left( \tfrac{n}{N}, N \right) 
  \, \phi \left( \tfrac{n}{N} \right)
  = \int_{\R} K_2^{\rm CSE} (\tau) \, \phi(\tau) \, \ud\tau ,
\end{equation}
with \cite{Mehta}
\begin{equation}
\label{KCSE}
  K_2^{\rm CSE}(\tau)
  = \begin{cases} \frac{1}{2} |\tau| - \frac{1}{4}|\tau| \log
    \abs{1 - \tau} & \text{for $|\tau| \le 2$} \\
    1  & \text{for $|\tau| > 2$} \, .
\end{cases}
\end{equation}
This statement implies that also the pair correlation
density $R_2(s,N)$ converges  (on average) to the
corresponding CSE density.  

It seems to be extremely difficult to prove rel.~(\ref{theBGS}) with
present techniques. Here, we will aim at understanding  
the asymptotics of
the form factor for small values of $\tau$, as $N\rightarrow\infty$,
in the regime governed by the {\em diagonal approximation}
\cite{HanOzo84,Ber85}.  For the diagonal approximation to
hold, we will assume in the following that there are no systematic
degeneracies in the classical actions 
of the map under consideration. This is true for generic
perturbed cat maps  $\vecphi$, but not for the original 
cat maps $A$, where
actions are in fact highly degenerate \cite{Kea91a,Kea91b}.

By virtue of our trace formula \eqref{trform} we have, for 
$\tau=\frac{n}{N}\neq 0$, $N\rightarrow \infty$,%
\begin{equation}
\label{scff}
K_2(\tau,N)=
\frac{1}{4N} \abs{\uTr \left(\opU^n
  \right)}^2 \sim \frac{1}{4N} \sum_{\bzf, \bzf^{\prime}} \frac{\utr
  d_n(\bzf) \utr d_n(\bzf^{\prime})}{\sqrt{-R^{\left(n\right)}_{\bzf}}
  \sqrt{-R^{\left(n\right)}_{\bzf^{\prime}}}}\exp \left [2 \pi \ui N
  \left(S_{\vecphi^n}(\bzf) - S_{\vecphi^n}(\bzf^{\prime}) \right) \right]
\end{equation}
(recall $\utr d_n(\bzf^{\prime})=\overline{\utr d_n(\bzf^{\prime})}$
for $d_n\in \SU(2)$),
where the sum extends over all fixed points of order $n$.
%
%\begin{equation}
%S_{\vecphi^n}(\bzf)=
%S_{\vecphi^n}
%(q_{\text{f}} + m_1,q_{\text{f}}) - m_2q_{\text{f}}
%\end{equation} 
%
%is the classical action. 
%
 Since for
$S_{\vecphi^n}(\bzf) \neq S_{\vecphi^n}(\bzf^\prime)$, cf. (\ref{action}),
 the exponential in \eqref{scff} shows rapid oscillations as 
$N \to \infty$, we assume that in the combined limit 
\begin{equation}
\label{limit}        
\tau\to 0 \, , \quad N\to\infty \, , \quad  n = \tau N \to \infty  
\end{equation}
the double sum in \eqref{scff} is dominated by the 
diagonal terms \cite{HanOzo84,Ber85}, i.e. by the terms with 
$S_{\vecphi^n}(\bzf)=S_{\vecphi^n}(\bzf^\prime)$. 
We know that 
\begin{equation}
S_{\vecphi^n}(\bzf)=S_{\vecphi^n}(\bzf^\prime) ,\quad
R^{\left(n\right)}_{\bzf}=R^{\left(n\right)}_{\bzf^\prime}, \quad
\utr d_n(\bzf)=\utr d_n(\bzf^\prime)
\end{equation}
for all points $\bzf^\prime$ along the periodic orbit
(compare section~\ref{stf}),
\begin{equation}
\bzf^\prime = \vecphi(\bzf), \ldots, \vecphi^{n-1}(\bzf), 
\vecphi^n(\bzf)=\bzf.
\end{equation}
Thus, we find $n^\#$ degenerate actions for every orbit of period $n$,
$n^\#$ denoting the {\it primitive} period, i.e. 
 $n=k n^\#$, $k\in \mathbb{N}$.
 Furthermore, since our system is invariant under 
(non-conventional) time-reversal $\tilde{\opT}$, we have 
\begin{equation}
S_{\vecphi^n}(\bzf)=S_{\vecphi^{n}}(\tilde{\vectau}(\bzf)) ,\quad
R^{\left(n\right)}_{\bzf}=R^{\left(n\right)}_{\tilde{\vectau}(\bzf)} , \quad
\utr d_n(\bzf)=\utr d_{n}(\tilde{\vectau}(\bzf)) .
\end{equation}
 Neglecting self-retracing orbits in the limit $n \to \infty$  
this results in an additional factor of 2 in the diagonal 
approximation. Note that if the classical map $\vecphi$ has a further
symmetry, e.g. inversion $\vecpi$ (\ref{inversion}), we also have 
$S_{\vecphi^n}(\vecpi(\bzf))=S_{\vecphi^{n}}(\bzf)$. However, since 
in general $\utr d_n(\bzf)$ and $\utr d_n(\vecpi(\bzf))$ become 
uncorrelated for large $n$, we will neglect cross correlations of these 
terms. We therefore conclude that in the 
combined limit (\ref{limit}) the form factor can be approximated by  
\begin{equation}
\label{asymptc}
  K_2(\tau,N) 
  \sim \frac{2n}{4N} \sum_{\bzf}  \frac{\left(\utr d_n(\bzf)
  \right)^2}{-R^{\left(n\right)}_{\bzf}}  .
\end{equation}

It is well known \cite{ParPol90}  
that the number of  fixed points of Anosov maps
grows, on average, like $\sim \ue^{hn}$ for $n$ large,
where $h$ denotes the topological entropy.
What is more, the fixed points become equidistributed in phase space $\T^2$
\cite{HanOzo84,ParPol90}, in the sense that 
for any smooth test function $a(\vecz)$, we have for $n$ large
\begin{equation}\label{sumrule}
\bigg\langle \sum_{\bzf}  \frac{a(\bzf)}
{-R^{\left(n\right)}_{\bzf}}\bigg\rangle_n 
\sim \int_{\T^2} a(\vecz) \, \ud z  ,
\end{equation}
where the average $\langle\ldots\rangle_n$ is some linear mean
over an interval about $n$ whose size grows to $\infty$ 
as $n\rightarrow\infty$. 
Let us ignore the fact that $d_n(\vecz)$ depends on $n$,
since, as we shall justify below, it in fact converges on average
to a constant independent of $n$. Hence by virtue of (\ref{sumrule}) 
\begin{equation}\label{all}
\bigg\langle \sum_{\bzf}  \frac{\left(\utr d_n(\bzf)
  \right)^2}{-R^{(n)}_{\bzf}} \bigg\rangle_n 
\sim \int_{\T^2}  \Big\langle  
\left(\utr d_n(\vecz)\right)^2 \Big\rangle_n \ud z  .
\end{equation}

In the semiclassical limit the quantum dynamics reduces 
to the skew product dynamical system  \cite{BolKep99b}
\begin{equation}
\begin{split}   
\bY : \
   \T^2 \times \SU(2) &\rightarrow \T^2 \times \SU(2) \\
   (\vecz,g) \quad & \mapsto \ (\vecphi(\vecz), \, g_1(\vecz) \, g) 
\end{split}     
\end{equation}
with $g_1(\vecz)=\exp \left(\ui \vecsig\cdot \vecB(q) \right)$. 
The $n$th iterate is then given by 
\begin{equation}
\bY^n(\vecz,g) = (\vecphi^n(\vecz), \, d_n(\vecz) \, g).
\end{equation}
The right-hand-side of (\ref{all}) can be viewed as a 
sequence of probability measures
\begin{equation}
\nu_n(F)= \int_{\T^2}  \Big\langle  
F(\bY^n(\vecz,\eins_2)) \Big\rangle_n \ud z  ,
\end{equation}
if we put $F(\vecz,g)=(\utr g)^2$.
For $F$ bounded on $\T^2\times\SU(2)$, this sequence is
contained in a compact space of probability measures,
hence we find a convergent subsequence $n_j$
with some limit measure $\mu$, i.e.
\begin{equation}
\lim_{j\rightarrow\infty} \nu_{n_j}(F) = \mu(F) .
\end{equation}
Due to the $\langle\ldots\rangle_n$ average we have,
for $n$ large,
\begin{equation}
\nu_{n}(F\circ\bY^r) \sim \nu_n(F) ,
\end{equation}
for any {\em fixed} integer $r$.
 Moreover, (with the substitution $w=\vecphi(z)$) we have 
\begin{equation}
\begin{split}   
  \nu_n(F \circ \bY) 
  & = \int_{\T^2} \Big\langle F \left( \vecphi^{n+1}(z), \, d_{n+1}(z) \right) 
                  \Big\rangle_n \, \ud z
    = \int_{\T^2} \Big\langle F \left( \vecphi^{n}(w), \, 
                  d_{n+1}(\vecphi^{-1}(w)) \right) \Big\rangle_n \, \ud w
  \\
  & = \int_{\T^2} \Big\langle F \left( \vecphi^{n}(w), \, d_{n}(w) 
                        \, g_1(\vecphi^{-1}(w)) \right) 
                  \Big\rangle_n \, \ud w
    = \nu_n(F\circ \vecgam_1)
\end{split}     
\end{equation}
where $\vecgam_1$ is defined by 
\begin{equation}        
  \gamma_1: (z,g) \mapsto (z, \, g \, g_1(\phi^{-1}(z))) \, .
\end{equation}
Therefore, we have $\nu_{n}(F\circ\vecgam_1) \sim \nu_n(F)$, 
and similarly, since
\begin{equation}
\int_{\T^2} F \left( \vecphi^{n+r}(\vecz), \, d_{n+r}(\vecz) \right) \, \ud z
= \int_{\T^2} F \left( \vecphi^n(\vecz), \, d_n(\vecz) 
                       \, d_r(\vecphi^{-r}(\vecz)) \right) \, \ud z,
\end{equation}
for any fixed $r \in \Z$ we obtain
\begin{equation}
  \nu_{n}(F\circ\vecgam_r) \sim \nu_{n}(F),
\end{equation}  
with $\vecgam_r: (\vecz,g) \mapsto 
      \left( \vecz, \, g \, d_r(\vecphi^{-r}(\vecz)) \right)$. 
Thus, we conclude that the limit measure $\mu$ itself must be 
invariant, i.e. $\mu(F\circ\tilde{\vecgam}_r) = \mu(F)$, for all $r\in\Z$.
Clearly, since $\mu$ is invariant under $\vecgam_r$, 
it is also invariant under $\vecgam_r^{-1}$ 
and in particular we obtain the relation
\begin{equation}
  \mu(F\circ\tilde{\vecgam}_r) \sim \mu(F) 
  \quad {\rm with} \quad
  \tilde{\vecgam}_r := \vecgam_r \circ \vecgam_{r-1}^{-1} : 
  (z,g) \mapsto \left( z, \, g \, g_1(\phi^{-r}(z)) \right) . 
\end{equation}  

So if, for almost all $\vecz\in\T^2$,
we find a set $V$ of integers such that the group generated by 
$\{g_1(\vecphi^{-\nu}(\vecz))\}_{\nu\in V}$ is dense in $\SU(2)$, 
then the action of that group is uniquely ergodic on $\SU(2)$
(see \cite{GamJakSar99} for more details and references
on equidistribution on $\SU(2)$), and we have that
$d\mu=dz\, dg$, where $dg$ denotes Haar measure. 
Clearly it is always possible to satisfy the above condition
for any generic choice of $g_1$, i.e. for any
generic magnetic field $\vecB(q)$, cf. section \ref{symmetry}. 

Since the limit $\mu$ of every converging subsequence is unique,
in fact {\em every} subsequence converges to $\mu$. That is
\begin{equation}\label{yall}
\lim_{n\rightarrow\infty}  \int_{\T^2}  \bigg\langle  
F(\bY^n(\vecz,\eins_2)) \bigg\rangle_n \ud z  
= \int_{\T^2}\int_{\SU(2)} F(\vecz,g) \, \ud z \, \ud g .
\end{equation}
In our case $F(\vecz,g)=(\utr g)^2$, and by the character formula
\cite{Ser77}
\begin{equation}\label{su2}
\int_{\T^2}\int_{\SU(2)} F(\vecz,g) \, \ud z \, \ud g =
\int_{\SU(2)} (\utr g)^2 \, \ud g = 
1 .
\end{equation}

Therefore, for generic magnetic fields $\vecB(q)$, 
the asymptotics of the form factor at small $\tau$
is in the diagonal approximation given by 
\begin{equation}
  K_2(\tau,N) \sim \frac{1}{2} \tau \, ,
\end{equation}
which is identical to the small $\tau$ asymptotics of the CSE form factor
\eqref{KCSE}.

\section{Piecewise constant magnetic fields}\label{piecewise}

As we have seen in the previous section, the diagonal approximation works
when the spin precession becomes equidistributed in $\SU(2)$. 
It is quite remarkable that this is not a necessary condition.
Suppose for instance that the magnetic field is piecewise constant,
such that $g_1(\vecphi^1(\vecz))$, $g_1(\vecphi^2(\vecz))$,\ldots 
only takes values in a discrete set in $\SU(2)$. In this case
the equidistribution theorem (\ref{yall}) holds with 
$\SU(2)$ replaced by $\Gamma$, where $\Gamma$ is the finite 
subgroup generated by $\{ g_1(\vecphi^r\vecz)\}_r$,
\begin{equation}\label{yall2}
 \lim_{n\rightarrow\infty}  \int_{\T^2}  \bigg\langle  
F(\bY^n(\vecz,\eins_2)) \bigg\rangle_n \ud z  
 = \frac{1}{\abs{\Gamma}} \sum_{g\in\Gamma} 
\int_{\T^2} F(\vecz,g) \, \ud z,
\end{equation}
 where $\abs{\Gamma}$ is the order of $\Gamma$. Now
\begin{equation}
   K_2(\tau,N) \sim \frac{1}{2} \tau \frac{1}{\abs{\Gamma}}
\sum_{g\in\Gamma} (\utr g)^2 .
\end{equation}
If the representation of $\Gamma$ in $\SU(2)$ is irreducible,
we have as a consequence of Schur's Lemma (see, e.g., 
\cite{Ser77} for details)
\begin{equation}
 \frac{1}{\abs{\Gamma}} \sum_{g\in\Gamma} (\utr g)^2 =1 ,
\end{equation}
as above. Irreducibility is guaranteed by our assumption in section
\ref{symmetry} 
that there be no vector $\vecr\neq0$ such that $\vecr\cdot\vecB(q)=0$
for all $q$.  
Hence we have again
\begin{equation}
  K_2(\tau,N) \sim \frac{1}{2} \tau \sim K_2^{\rm CSE}(\tau) ,
\end{equation}
for $\tau$ small.

\begin{figure}
\centering \subfigure[Spacing distribution of $\opU(A)$ as defined
  in~\eqref{maptd}]{
  \includegraphics[width=2.9in]{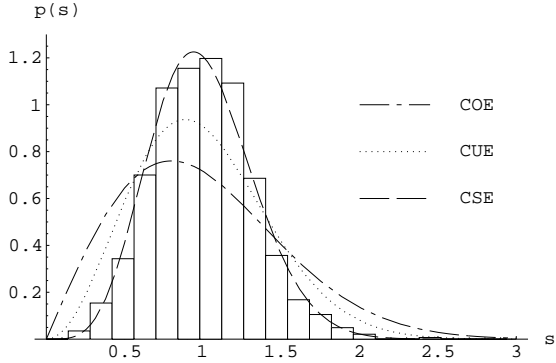}} \hspace{.2in}
\subfigure[Cumulative spacing distribution of $\opU(A)$ as defined
  in~\eqref{maptd}]{
  \includegraphics[width=2.9in]{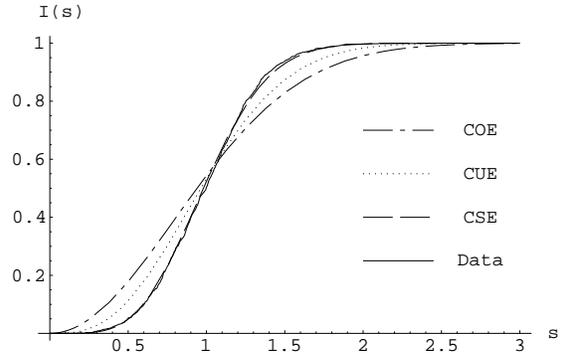}}\\
\subfigure[Spacing distribution of $\opU(\vecphi)$ as defined
  in~\eqref{maptd}.  The perturbation parameter is  k=0.32.]{
  \includegraphics[width=2.9in]{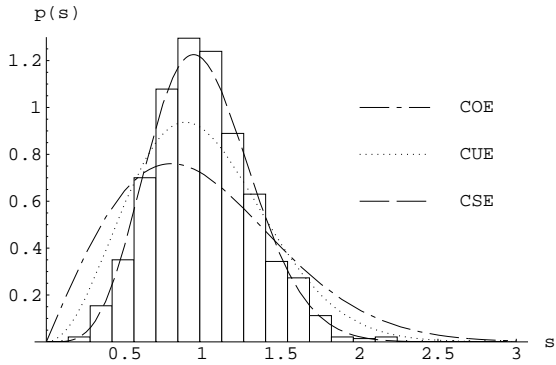}} \hspace{.2in}
\subfigure[Cumulative spacing distribution of $\opU(\vecphi)$ as defined with
k=0.32.]{ \includegraphics[width=2.9in]{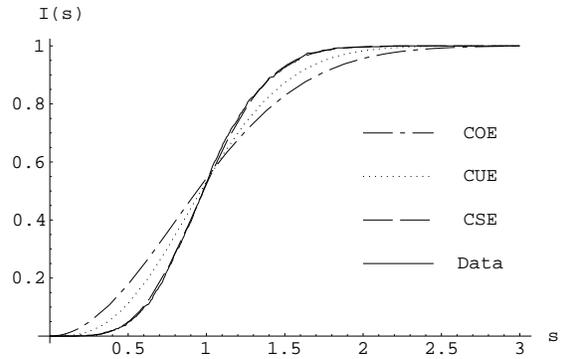}}
      \caption{Local spectral statistics  of the quantum cat
        map (above) and of its perturbation (below) when the spin
        precession is caused by the magnetic field~\eqref{mft}. 
        The number of eigenangles $N$ is
        $1021$.}
        \label{fig3}
\end{figure}

Let us complement this argument by a numerical experiment. 
We choose the magnetic field
\begin{equation}
\label{mft}
\vecB(q)=
\begin{cases}
 (\pi,0,0) & \text{if $0 \le q < 1/6$}  \\
 (0,\pi,0) & \text{if $1/6 \le q < 1/3$} \\
 (0,0,\pi) & \text{if $1/3 \le q < 1/2$} \\
    (0,0,0) & \text{if $q=1/2$.}
\end{cases} 
\end{equation}
The symmetry constraint $\vecB(1-q)=-\vecB(q)$ determines the
magnetic field in the interval $1/2 < q < 1$. We also require
periodicity, i.e. $\vecB(q + m) = \vecB(q)$, $ m \in \Z$.
This particular choice leads to the discrete group of Hamilton's quaternions,
\begin{equation}
\Gamma=\{ \pm\eins_2,\pm \ui \sigma_x,\pm \ui \sigma_y,\pm \ui\sigma_z\}.
\end{equation}
The quantum maps $\opU(A)$ and $\opU(\vecphi)$ defined
in~\eqref{maptd} can also be easily diagonalized with the magnetic 
field given by~\eqref{mft}.  The statistics are clearly CSE as
shown in fig.~\ref{fig3}. 
This outcome could be expected for
generic Anosov maps, since the diagonal approximation for the form factor
agrees, at small argument, with the random matrix prediction.

In the case of the unperturbed cat map, however, we cannot proceed as 
in section \ref{formf} due to the exponentially large number of 
fixed points sharing the same classical action \cite{PerViv87,Kea91a}. 
Remarkably, we still observe CSE statistics (fig. \ref{fig3}). 
Similar peculiarities are found
for arithmetic triangles with non-conventional boundary
conditions \cite{AurSchSte95}, non-arithmetic Hecke triangles 
\cite{BogGeoGiaSch97} and hyperbolic tetrahedra \cite{AurMar96}.

\section*{Acknowledgment}
We gratefully acknowledge  helpful discussions with Jens~Bolte,
Grischa~Haag, Jon Keating and Jonathan Robbins. S~K was partly supported by Deutscher Akademischer 
Austauschdienst (DAAD) under grant no. D/99/02553 and by Deutsche 
Forschungsgemeinschaft (DFG) under contract no. STE~241/10-1.  F~M 
was supported by a Royal Society Dorothy Hodgkin Fellowship 
during the period when this research was completed.

\appendix

\section{Quantum mechanics on the torus}
\label{qmt}

We briefly review  quantum mechanics of systems 
whose classical phase space is the torus $\T^2$. 
For more details see \cite{HanBer80,Esp93,EspGraIso95}.

Because of the topology of the torus, quantum states are taken to be
periodic in both  position and momentum representation, i.e. 
\begin{equation}
\psi(q + m_1) =  \psi(q), \qquad
\hat{\psi}(p + m_2) = \hat \psi(p), \qquad m_1,m_2\in\Z
\end{equation}
where
\begin{equation} \label{FT}
\hat{\psi}(p)=  \frac{1}{\sqrt{2\pi
    \hbar}}\int_{-\infty}^{+\infty} \psi(q)\ue^{-\frac{\ui}{\hbar}qp} \, \ud q.
\end{equation}
The periodicity of the wavefunction in both bases has two important 
consequences. Firstly, both $\psi(q)$ and $\hat\psi(p)$ are 
superpositions of delta functions supported on the lattices points 
$q=2\pi\hbar \; Q$ and $p=2 \pi \hbar\; P$ respectively, where $Q,P
\in \Z$, i.e. 
\begin{equation}
\psi(q)=\sum_{m \in \Z} \sum_{Q=0}^{N-1} \Psi(Q) \, 
\delta\left(q-\frac{Q}{N}+m\right) 
\end{equation}
with $\Psi(Q +N)=\Psi(Q)$. 
Secondly, $2 \pi \hbar$ must be an inverse integer,
i.e. $N=1/2\pi\hbar$.  It follows that the Hilbert space may be
identified with the $N$-dimensional vector space $\mathcal{H}_N$ with
inner product
\begin{equation}
(\Phi,\Psi)= \frac1N \sum_{Q \bmod N} \overline\Phi(Q) \Psi(Q). 
\end{equation}

In order to quantize observables $f\in C^\infty(\T^2)$
we need to introduce the translation operators
\begin{subequations}
\begin{align}
t_1 \Phi(Q)& = \Phi(Q+1) \\
t_2 \Phi(Q) &= \Phi(Q) \exp \left(\frac{2 \pi \ui}{N}\, Q\right),
\end{align}
\end{subequations} 
which may be viewed as the exponentials of the usual differentiation
and multiplication operators on the real line.
For any $m,n \in \Z$, we have the following commutation
relation
\begin{equation}
t_1^m t_2^n= \exp \left(\frac{2 \pi \ui}{N}\, mn \right)t_2^nt_1^m .
\end{equation}
Note that $t_1^N=t_2^N=\eins_N$.
The  Weyl-Heisenberg operators are defined by 
\begin{equation}
\label{heisop}
\OpTH(\bn)= \exp \left(\frac{\pi \ui}{N}\,n_1n_2 \right)t_2^{n_2}t_1^{n_1},
\end{equation}
where $\bn=(n_1,n_2)$.  We then have the following multiplication rule
\begin{equation}
\label{mrule}
\OpTH(\bm)\OpTH(\bn)= \exp\left(\frac{\pi \ui}{N}\,\vp{\bm}{\bn}\right)
\OpTH(\bm + \bn),
\end{equation}
where $\vp{\bm}{\bn}= m_1n_2-m_2n_1$ is the standard symplectic form.

Let $f \in C^{\infty}(\T^2)$ be a classical observable on $\T^2$ whose
Fourier series is given by
\begin{equation}
f(\bz) = \sum_{\bm \in \mathbb{Z}^2} \hat f_{\bm}\ue^{2 \pi \ui
  \bz\cdot\bm}, \quad \bz=(p,q) \in \T^2. 
\end{equation}
The Weyl quantization of $f$ is defined as
\begin{equation}
\label{WqT}
\Op(f)= \sum_{\bm \in \Z^2} \hat{f}_{\bm} \OpTH(\bm) \, .
\end{equation} 
Since $\Op(f^*)=\Op(f)^{\dagger}$, $\Op(f)$ is Hermitian if and only if
$f(\bz)$ is real.

Semiclassically,  quantum observables commute,
 which can be easily seen by expanding the 
exponential in (\ref{mrule}).  More precisely, 
let $f,g \in C^{\infty}(\T^2)$, then 
\begin{equation}
\label{semprop}
\Op(f)\Op(g) \sim \Op(fg), \quad N \rightarrow \infty.
\end{equation}
If $f$ and $g$ depend only on  either $p$ or $q$, the above
relation is an identity for each $N$.  The quantization of classical
matrix-valued observables is analogous to the case of scalar
functions.  Let $g(\bz)$ be a $2 \times 2$ matrix such that $g_{jk}
\in C^{\infty}(\T^2)$. We define its Weyl quantization as
\begin{equation}
\Op(g) = \begin{pmatrix} \Op(g_{11}) & \Op(g_{12}) \\
                         \Op(g_{21}) & \Op(g_{22}).
         \end{pmatrix} 
\end{equation}
From~\eqref{semprop} it follows that
\begin{equation}
\label{prrule}
 \Op(g_k\ldots g_2g_1) \sim \Op(g_k)\ldots \Op(g_2)\Op(g_1), \quad N
\rightarrow \infty.
\end{equation}
Clearly, we have
\begin{equation}
\label{hcom}
\Op(g^{\dagger})=\Op(g)^{\dagger}.
\end{equation}
If $g(\bz)$ is unitary, combining~\eqref{prrule} and~\eqref{hcom} yields
\begin{equation}
\Op(g)\Op(g)^{\dagger} \sim \eins_{2N}, \quad N \rightarrow \infty,
\end{equation}
that is $\Op(g)$ is only semiclassically unitary.  

We can now formulate a version of Egorov's theorem \cite{BouDeB96}
 for matrix valued observables. Let
\begin{equation}
 \opUt(\vecphi) = 
\begin{pmatrix} \opU_N(\vecphi) & 0 \\ 0 & \opU_N(\vecphi)\end{pmatrix},
\end{equation}
where $\opU_N(\vecphi)$ is the quantum propagator of an Anosov map
$\vecphi$.  We have
\begin{equation}
\label{egth}
\begin{split}
 \opUt^{-1}(\vecphi) \Op(g)\opUt(\vecphi)& = \begin{pmatrix} 
 \opU_N(\vecphi)^{-1}\Op(g_{11})\opU_N(\vecphi)&
 \opU_N(\vecphi)^{-1}\Op(g_{12})\opU_N(\vecphi) \\
 \opU_N(\vecphi)^{-1}\Op(g_{21})\opU_N(\vecphi) &
\opU_N(\vecphi)^{-1}\Op(g_{22})\opU_N(\vecphi) 
\end{pmatrix} \\
& \sim \begin{pmatrix} 
 \Op(g_{11}\circ \vecphi ) &
 \Op(g_{12} \circ \vecphi ) \\
 \Op(g_{21} \circ \vecphi ) &
\Op(g_{22} \circ \vecphi)
\end{pmatrix}= \Op(g \circ \vecphi), 
\end{split}
\end{equation}
as $N \rightarrow \infty$. 
For cat maps  this relation is an identity.

\section{Trace formula for matrix elements}
\label{trss} 

The following trace formula for matrix elements 
of quantum observables is a well known generalization
of Gutzwiller's trace formula \cite{Wil87,PauUri95,Bol00}. 
We will here present its derivation in the case of
general Anosov maps $\vecphi$ on the torus $\T^2$,
which is particularly clean and simple. 
The formula is needed in section \ref{stf}.

Suppose $\varphi_1,\ldots,\varphi_N$ is an orthogonal basis of
eigenfunctions of $\opU_N(\vecphi)$ with eigenphases $\omega_j$,
\begin{equation}
\opU_N(\vecphi) \varphi_j=\ue^{\ui \omega_j}\varphi_j .
\end{equation}

\noindent  
{\em Claim:} For any smooth 
classical observable $f\in\C^\infty(\T^2)$ and $n$ fixed we have
in the semiclassical limit ($N\rightarrow\infty$)
\begin{equation}\label{matrformapp}
\sum_{j=1}^N (\varphi_j, \Op(f) \varphi_j)\, \ue^{\ui n \omega_j}
\sim
\sum_{\bzf}
 \frac{f(\bzf)}{\sqrt{-R^{\left(n\right)}_{\bzf}}}
\exp\left[2 \pi \ui N\left( S_{\vecphi^n}
(q_{\text{f}} + m_1,q_{\text{f}}) - m_2q_{\text{f}}\right)
\right] ,
\end{equation}
where $(m_1,m_2)=\bm $ are the winding numbers and  $\bzf$ 
are the  fixed points of order $n$, i.e.
\begin{equation}
\vecphi^n(\bzf)=\bzf + \bm.
\end{equation}
Furthermore, $R^{\left(n\right)}_{\bzf}= \det
(\mathcal{M}^n_{\bzf}- I)$, where 
$\mathcal{M}^n_{\bzf} = \frac{\ud \vecphi}{\ud \bz}(\bzf)$ denotes the
monodromy matrix and $S_{\vecphi^n}$ generates the classical map $\vecphi^n$.

The left-hand-side of (\ref{matrformapp}) is of course the trace
of $\opU_N^n(\vecphi)\Op(f)$. Since
\begin{equation}
\Op(f)= \sum_{\vecmu \in \Z^2} \hat{f}_{\vecmu} \OpTH(\vecmu) \, , 
\end{equation} 
and because $f \in C^\infty(\T^2)$ implies that 
the coefficients $\hat{f}_{\vecmu}$ are rapidly decreasing,
we may without loss of generality consider only the  
matrix elements $\opU_N^n(\vecphi)\OpTH(\vecmu)$.
For $\nu=0,\ldots,N-1$  the functions
\begin{equation}
\delta_\nu(Q) = 
\begin{cases} 
\sqrt N  & (Q=\nu \bmod N) \\
0  & (Q\neq\nu \bmod N)
\end{cases}
\end{equation}
form an orthonormal basis of the  Hilbert space $\mathcal{H}_N$.
Hence we may write
\begin{equation} \label{trrr}
\begin{split}
\uTr \left[\opU_N^n(\vecphi)\OpTH(\vecmu) \right]
& = \sum_{\nu\bmod N} (\delta_\nu, \opU_N^n(\vecphi) 
\OpTH(\vecmu)\delta_\nu) \\
& =\sum_{\nu\bmod N} 
\exp \left(\frac{\pi \ui}{N}\,\mu_1\mu_2 \right) 
\exp \left(\frac{2\pi \ui}{N}\,\mu_2 (\nu-\mu_1) \right) 
(\delta_\nu, \opU_N^n(\vecphi) \delta_{\nu-\mu_1}) \\
& \sim \sum_{\nu\bmod N} 
\exp \left(\frac{2\pi \ui}{N}\,\mu_2 \nu \right) 
(\delta_\nu, \opU_N^n(\vecphi) \delta_{\nu-\mu_1}) 
\end{split}
\end{equation}
for $N$ large and  $\|\vecmu\| < N^{1/2-\delta}$, $\delta >0$.  
As we shall justify below any smooth observable can be well approximated 
with a suitable cut-off in $\vecmu$. 
In the semiclassical limit the matrix elements of $\opU_N^n(\vecphi)$ 
are given by \cite{BdMOdA95}  
\begin{equation}
\label{mels}
(\delta_\nu, \opU_N^n(\vecphi) 
\delta_{\nu'})\sim \frac{1}{M_n}\sum_{m_1=0}^{M_n} 
D \left( \frac{\nu}{N} + m_1,\frac{\nu'}{N} \right)
\exp\left[2\pi\ui NS_{\vecphi^n} 
  \left( \frac{\nu}{N} + m_1, \frac{\nu'}{N} \right) \right] , 
\end{equation}
where 
\begin{equation}
D(q_2,q_1)= \left( \frac{ \ui }{N} \frac{\partial^2
    S_{\vecphi^n}(q_2,q_1)}{\partial q_2 \partial q_1} \right)^{1/2},
\end{equation}
$S_{\vecphi^n}(q_2,q_1)$ is the classical action of the lift of 
$\vecphi^n$ on the covering plane~\cite{KeaMez00}, and $M_n$ depends
on the unperturbed map and on $n$.

Inserting~\eqref{mels}
into~\eqref{trrr} and applying the Poisson summation formula yields
\begin{multline}
\label{t4r}
\uTr \left[\opU_N^n(\vecphi)\OpTH(\vecmu) \right]
 \sim \frac{1}{M_n}\sum_{m_1=0}^{M_n} \sum_{m_2 \in \Z} N
  \int_{-\epsilon}^{1-\epsilon}  D \left( q + m_1,q-\frac{\mu_1}{N} \right) 
  \times \\ \times 
  \exp\left[2\pi\ui N \left(
    S_{\vecphi^n} \left( q + m_1, q-\frac{\mu_1}{N} \right) -m_2 q \right)
  \right] \exp \left(2\pi \ui\,\mu_2 q \right) \ud q 
\end{multline}
In leading order ($N \rightarrow \infty$) we have,
uniformly for all $\vecmu$ with $\|\vecmu\| \leq N^{1/2-\delta}$,  
\begin{equation}
\label{semap}
  D \left( q + m_1,q - \frac{\mu_1}{N} \right) 
  \sim D(q + m_1,q) 
\end{equation}
and
\begin{equation}
\begin{split}
S_{\vecphi^n} \left( q + m_1,q - \frac{\mu_1}{N} \right)& 
\sim S_{\vecphi^n}(q+ m_1,q) - 
\left. \frac{\partial S_{\vecphi^n}(q+ m_1,q')}{\partial
  q'}\right \rvert_{q'=q}\frac{\mu_1}{N}\\
& \sim  S_{\vecphi^n}(q+ m_1,q) +  p(q)\frac{\mu_1}{N} 
\end{split}
\end{equation}
Inserting~\eqref{semap} into ~\eqref{t4r} gives
\begin{multline}
\label{t4r2}
\uTr \left[\opU_N^n(\vecphi)\OpTH(\vecmu) \right] \sim
 \frac{1}{M_n} \sum_{m_1=0}^{M_n}
\sum_{m_2 \in \Z} N \int_{-\epsilon}^{1-\epsilon} 
    \exp \left[2\pi \ui\,(\mu_1 p(q) + \mu_2 q) \right] \times \\ \times
    D(q + m_1, q)  \exp \left[2 \pi \ui N\left( S_{\vecphi^n}(q +
        m_1, q) - m_2q\right) \right] \ud q . 
\end{multline}

The above expression is the same as the formula that we would have
obtained in determining the trace of the $n$-th propagator of an
Anosov map with $D_n(q_1,q_2)$ replaced by 
$D_n(q_1,q_2) \exp(2\pi \ui \vecmu\cdot \vecz)$.
Following the same steps which lead to the trace formula of Anosov
maps~\cite{BdMOdA95,KeaMez00}  we obtain
\begin{equation}\label{fine}
\uTr \left[\opU_N^n(\vecphi)\OpTH(\vecmu) \right]\sim
\sum_{\bzf}
 \frac{\exp(2\pi \ui \vecmu\cdot \bzf)}{\sqrt{-R^{\left(n\right)}_{\bzf}}}
\exp\left[2 \pi \ui N\left( S_{\vecphi^n}
(q_{\text{f}} + m_1,q_{\text{f}}) - m_2q_{\text{f}}\right)
\right] .
\end{equation}
Let $f_N$ be a truncated Fourier approximation to $f$,
\begin{equation}
f_N(\bz)= \sum_{\substack{\vecmu \in \Z^2 \\ \|\vecmu\| < N^{1/2-\delta}}} 
\hat{f}_{\vecmu}  
\exp(2\pi \ui \vecmu\cdot \bz) . 
\end{equation} 
Taking finite linear combinations, (\ref{fine}) yields
\begin{equation}\label{matrformapp2}
\sum_{j=1}^N (\varphi_j, \Op(f_N) \varphi_j)\, \ue^{\ui n \omega_j}
\sim
\sum_{\bzf}
 \frac{f_N(\bzf)}{\sqrt{-R^{\left(n\right)}_{\bzf}}}
\exp\left[2 \pi \ui N\left( S_{\vecphi^n}
(q_{\text{f}} + m_1,q_{\text{f}}) - m_2q_{\text{f}}\right)
\right] .
\end{equation}
To conclude the proof of  our claim (\ref{matrformapp}), we note
that, since the Fourier coefficients of $f$ are decreasing
faster than any power, both
\begin{equation}
\big|\sum_{j=1}^N (\varphi_j, \Op(f) \varphi_j)\, \ue^{\ui n \omega_j}
-(\varphi_j, \Op(f_N) \varphi_j)\, \ue^{\ui n \omega_j}\big|
\leq \sum_{j=1}^N \big|(\varphi_j, \Op(f-f_N) \varphi_j) |
\end{equation} 
and 
\begin{equation}
\big|\sum_{\bzf}
 \frac{f(\bzf)-f_N(\bzf)}{\sqrt{-R^{\left(n\right)}_{\bzf}}}
\exp\left[2 \pi \ui N\left( S_{\vecphi^n}
(q_{\text{f}} + m_1,q_{\text{f}}) - m_2q_{\text{f}}\right)
\right] \big|
\leq
\sum_{\bzf}
\big| \frac{f(\bzf)-f_N(\bzf)}{\sqrt{-R^{\left(n\right)}_{\bzf}}}\big|
\end{equation} 
are rapidly  decreasing, as $N$ becomes large.

\bibliographystyle{my_unsrt}
\bibliography{literatur}

\end{document}